\documentclass[aps,prl,twocolumn,superscriptaddress,amsmath,amssymb,showpacs]{revtex4}
\usepackage{graphicx}
\usepackage{dcolumn}
\usepackage{bm}
\begin{document}

\title{A wide energy-window view on the density of states and hole mobility in poly({\it p-}phenylene vinylene)}

 \author{I. N. Hulea}

 \affiliation{Kamerlingh Onnes Laboratory, Leiden University, POB. 9504, 2300 RA Leiden, The Netherlands}

 \affiliation{Dutch Polymer Institute (DPI), POB. 902, 5600 AX Eindhoven, The Netherlands}

 \author{H. B. Brom}

 \affiliation{Kamerlingh Onnes Laboratory, Leiden University, POB. 9504, 2300 RA Leiden, The Netherlands}

 \author{A. J. Houtepen}

 \author{D. Vanmaekelbergh}

 \author{J. J. Kelly}

 \affiliation{Debye Institute, Utrecht University, POB. 80000, 3508 TA Utrecht, The Netherlands}

 \author{E. A. Meulenkamp}

 \affiliation{Philips Research Laboratories, Prof. Holstlaan 4, 5656 AA Eindhoven, The Netherlands}

\date{September 7 2004 accepted for Phys. Rev. Letters}

\begin{abstract}
Using an electrochemically gated transistor, we achieved
controlled and reversible doping of poly({\it p-}phenylene
vinylene) in a large concentration range. Our data open a wide
energy-window view on the density of states (DOS) and show, for
the first time, that the core of the DOS function is Gaussian,
while the low-energy tail has a more complex structure. The hole
mobility increases by more than four orders of magnitude when the
electrochemical potential is scanned through the
DOS.\end{abstract}


 \maketitle

Charge transport in disordered conjugated polymers like poly({\it
p-}phenylene vinylene) (or PPV), polypyrrole and polythiophene)
\cite{Kaiser01}, and disordered systems in general
\cite{Kador91,Bassler93,Young95,Dong98,Roest00} usually proceeds
via thermally activated hopping allowing the charge carriers to
move from one site to the next. In this process the
energy-dependent density of states (DOS) and the charge mobility
($\mu$) (or the diffusion constant ($D$)) are two key parameters.
The energy distribution of the DOS is often assumed to be Gaussian
\cite{Bassler93,Hirao97,Novikov98} [for dipolar interaction
\cite{Martens000} the width will be proportional to the strength
of the dipoles \cite{Kador91,Young95,Novikov95}] or exponential
\cite{Vissenberg98} and both shapes have been applied with success
to explain transport properties under different conditions
\cite{Wang00,Schmechel02,Arkhipov03,Shaked03}. In devices such as
polymeric light emitting diodes (LEDs), where the carrier
concentration is low, only the tail of the DOS is directly
involved in the charge injection. In field effect transistors
(FETs) the carrier concentration c is orders of magnitude larger
and the DOS further towards the center of the level distribution
is important. These examples show that, for a proper understanding
of the electronic properties of such materials, a direct
experimental determination of the shape of the DOS function over a
large range of energy (or charge concentration) is essential.
Until now, only a few measurements have been reported.\\ One
attempt to determine the DOS (from now on only valence-states are
considered) in ${\rm OC_1C_{10}}$-PPV, the workhorse in light
emitting diodes, has used the temperature and concentration
dependence of the hopping conductivity with FeCl$_3$ as a dopant
\cite{Martens03}. In another approach it is assumed that the DOS
is Gaussian shaped. It then follows
\cite{Bassler93,Novikov98,Martens000,Martens00,Tanase03} that the
carrier mobility $\mu$ depends on temperature via ${\rm ln}\mu
\propto -(\sigma_d/k_{\rm B}T)^2$. From the experimental
determination of the mobility as a function of temperature the
width $\sigma_d$ of the distribution was
determined\cite{Martens00,Tanase03} [for a discussion of some of
the simplifying assumptions see
\onlinecite{Novikov98,Baranovskii01}]. Recently, concentration
dependent DOS and $\mu$ data were obtained in a FET and a LED
configuration \cite{Tanase03}. All of the above mentioned
experiments showed the mobility to be strongly concentration
dependent. In the charge carrier concentration range covered by
chemical doping with FeCl$_3$ the DOS increased linearly with $c$
\cite{Martens03}, while for the analysis of the FET and LED
experiments \cite{Martens00,Tanase03} an exponential and Gaussian
DOS were assumed, respectively. It can be concluded that there is
still no consensus on the DOS of PPV in a broad energy range.
\noindent

This letter reports on the DOS and conductivity of thin
spin-coated PPV films in a wide, well defined energy range. The
data are obtained using an Electrochemically Gated Transistor
(EGT), which was recently
developed\cite{Meulenkamp99,Jiang02,Roest02}. Using the impressive
energy range of the EGT, we are able to confirm the previous data
and to show for the first time that the core of the DOS function
is Gaussian, while the flank has an exponential energy dependence.
The low energy tail has a more complex structure. Conductance
measurements were performed over a wide carrier concentration
range and show that the hole mobility is strongly energy
dependent, as also seen in results obtained with a classical FET
\cite{Tanase03}. The maximum value of the mobility exceeds
previously reported values by a factor of 40. The mobility is
found to vary over more than 4 orders of magnitude when the
Fermi-level is scanned over less than half an eV. In addition, our
measurements allow an energy calibration of the DOS and the
mobility.

\begin{figure}[b]
\begin{center}
\includegraphics[width=6cm]{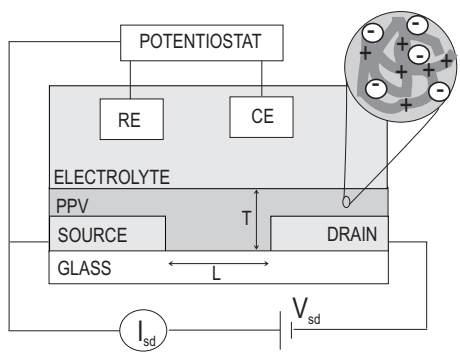}
\end{center}
\noindent{\caption{Schematic picture of the electrochemically
gated transistor or EGT. The sample (PPV) is placed in an
electrolyte solution. The electrochemical potential of the sample
is controlled with respect to a reference electrode (RE) using a
potentiostat. The conductivity was measured by applying a small dc
bias between source and drain. The enlargement shows schematically
how the holes on the polymer chain are compensated by the anions
in the film.} \label{Egt}}
\end{figure}

\begin{figure}[h]
\begin{center}
\includegraphics[width=7cm]{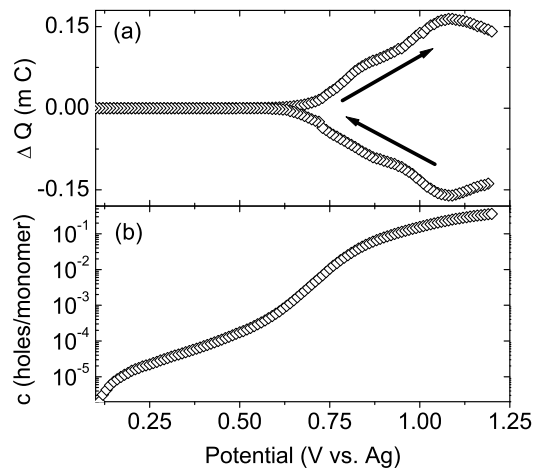}
\end{center}
\noindent{\caption{Electrochemical injection of holes into the PPV
film. (a) The differential charge $\Delta Q(\tilde{\mu_{e}})$
stored in PPV when successive 10 mV steps are applied. The
potential is defined with respect to the Ag pseudo-reference. The
arrows show the two scanning directions: doping and dedoping while
changing $\tilde{\mu_{e}}$. (b) Doping per monomer
$c(\tilde{\mu_{\rm e}})$ calculated from the data in (a). The
reversible doping range extends over 4 orders of magnitude.}
\label{dq}}
\end{figure}

The electrochemically gated transistor offers a unique possibility
of studying the transport properties of polymers as a function of
doping in a reversible way. The operating principle is based on
changing the electrochemical potential $\tilde{\mu_{\rm e}}$ of
the sample (${\rm OC_1C_{10}}$-PPV) with respect to a Ag
pseudo-reference electrode (RE) by means of a potentiostat, see
Fig.~\ref{Egt}. The PPV layer is in electrochemical equilibrium
with the Au source-drain electrodes. Any change in potential with
respect to the Ag electrode is followed by charge transfer from
the Au electrodes to the PPV or vice versa and current flow to the
Pt counter electrode (CE). When the potential is increased holes
are injected into the PPV \cite{noteelectrondope}. The hole charge
is counterbalanced by anions (ClO$^-_4$ or PF$^-_6$) from the
electrolyte solution (0.1 M TBAP (tetrabutylammonium perchlorate),
TBAPF$_6$ (tetrabutylammonium hexafluorophosphate) or LiClO$_4$ in
acetonitrile) \cite{chemicals} which permeates the PPV. The number
of holes stored in the PPV film is determined by monitoring the
differential capacitance. Important advantages of electrochemical
gating over the conventional field-effect transistor are the
uniform charging\cite{Roest02} of the PPV film together with a
wider doping range. At a given doping level the conductance is
measured with a very small source-drain bias (10mV, $I_{\rm SD}
\propto V_{\rm SD}$) supplied by a Keithley 2400 source meter.
Another advantage of the EGT is that the electrochemical potential
can be correlated with the vacuum level. To achieve this the
potential of the Ag pseudo-reference electrode was measured with
respect to the electrochemical potential of the
ferrocenium/ferrocene redox reaction (0.68~V). From the literature
value of the latter\cite{Noviandri99} with respect to vacuum
(5.14~V), the Ag pseudo-reference electrode was calculated
to be at 4.47~V below the vacuum level.\\
The PPV samples with a typical thickness $T$ of 180 nm were
spin-coated on glass substrates with Au electrodes separated by a
gap $L$ of 1.25 up to 10 $\mu {\rm m}$ and with an effective
transistor length of 1 up to 50 cm (interdigitated electrodes).
The electrochemical potential was controlled with a Princeton
Applied Research Potentiostat/Galvanostat 273A and all experiments
were carried out under Ar atmosphere. The volume $V_{\rm P}$ of
the various polymer films was between $0.03~\rm mm^3$ to $0.1~\rm
mm^3$ (accuracy 5\%). From the specific weight of the PPV one can
determine the density $N$ of PPV monomers \cite{notemonomer}. The
density of states $g(\tilde{\mu_{e}})$, was obtained from the
charge $\Delta Q$ stored in the PPV film per ${\Delta
\tilde{\mu_{e}}}=10$ meV increase of $\tilde{\mu_{e}}$, i.e.
 \begin{equation}
 g(\tilde{\mu_{e}}) = \frac{\Delta Q}{e \Delta \tilde{\mu_{e}}NV_P}
 \end{equation}
with $e$ the elementary charge (see Fig.~\ref{dq}a). The DOS is
therefore determined directly; no assumptions are required. From 0
to 0.6 eV the amount of injected charge is small. After this point
a strong increase is observed. At around 0.8 eV a shoulder is
visible and above 1.1 eV, $\Delta Q(\tilde{\mu_{e}})$ starts to
decrease. With decreasing $\tilde{\mu_{e}}$ the maximum and the
shoulder are well reproduced. Measurements on different samples,
with different potential steps and other ions in solution gave
very similar results. The number of holes per monomer
$c(\tilde{\mu_{e}})$ at a given electrochemical potential is
obtained by integration of $g(E)$:
$c(\tilde{\mu_{e}})=\int_{0}^{\tilde{\mu_{e}}}g(E)dE$ (see
Fig.~\ref{dq}b). It can be seen that the doping level can be
varied by electrochemical gating in a controlled and reversible
manner between $10^{-5}$ and 0.4 holes per monomer.

\begin{figure}[t]
\begin{center}
\includegraphics[width=8cm]{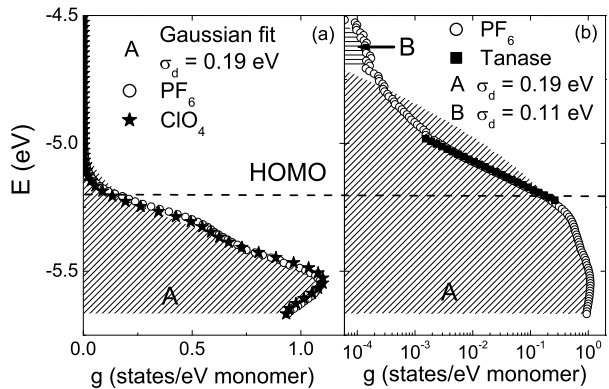}
\end{center}
\noindent{\caption{The $E$ dependence of the experimentally
determined DOS ($g(E)$) where $E$ is with respect to the vacuum
level. The horizontal dashed line marks the HOMO (Highest Occupied
Molecular Orbital) position found from cyclic voltammetry
\cite{vanDuren03,Hulea03}. (a) $E$ vs. $g(E)$ using PF$_6^-$ and
ClO$_4^-$ as anions. A Gaussian function with $\sigma =$ 0.19 eV
(area A) fits the data well. (b) $E$ vs. $g(E)$ for PF$_6^-$ on a
linear-log. scale. Below $g = 0.1\,{\rm states/(eV\,monomer)}$
deviations from the Gaussian fit are visible. The filled squares
are the PPV FET-data by Tanase {\it et al.} \cite{Tanase03}, which
are well described by an exponential function. At the lowest
values of $g(E)$ additional structure appears, which in a limited
energy range would allow a description with a Gaussian with a
width of 0.11 eV (area B) as used for PPV LEDs by Martens {\it et
al.} \cite{Martens000} and Tanase {\it et al.} \cite{Tanase03}
\label{densitE}}}
\end{figure}

The density of states g($E$) is plotted in Fig.~\ref{densitE},
where the energy on the vertical axis is given with respect to the
vacuum level. The values for $g(E)$ on the horizontal axis range
from $10^{-4}$ to 1 states/monomer eV. At $E$ = -5.55 $\pm$
0.02~eV, $g$ has a maximum. This shows that due to disorder and
dispersion the HOMO levels are spread in energy and we are able to
access the center of the distribution. The shoulder at about -5.3
eV reveals some additional structure in the measured DOS. By
fitting the data to a Gaussian distribution
 \begin{equation}\label{Gauss}
  g(E)= \frac{N_m}{\sqrt{2\pi}\sigma_d}\exp[-(\frac{E-E_{ct}}{\sqrt{2}\sigma_d})^2],
 \end{equation}
we find a width $\sigma_d=0.19 \pm~0.01$~eV, centered at $E_{\rm
ct}= -5.55\pm~0.02$~eV (diagonally shaded area A in
Fig.~\ref{densitE}). The number of states per monomer $N_m = \int
g(E)dE$ has a value of $0.52 \pm~0.01$ when we integrate over the
whole Gaussian \cite{Nm}. The measured DOS validates the
assumption of a Gaussian distribution \cite{Bassler93} in the core
($g>10^{-1}$ states$/$(eV monomer)). For $g$ between $10^{-3}$ and
$10^{-1}$ states$/$(eV monomer) the data follow an exponential
dependence, see the logarithmic-linear plot of
Fig.~\ref{densitE}b. The DOS in this range measured by Tanase {\it
et al.} in a solid state FET configuration \cite{Tanase03} (solid
squares in Fig.~\ref{densitE}b) is in very good agreement with our
data.

In general ions in the dielectric will influence the energy
landscape and broaden the DOS \cite{Arkhipov03}. The data
presented here compare well with those of Tanase et al.
\cite{Tanase03}, where no ions are present in the film. Hence, the
intrinsic level distribution of the investigated ${\rm
OC_1C_{10}}$-PPV must be sufficiently broad, such that an extra
contribution by the ions is not observable in our experiments.
There is no guarantee that this also holds for the data at higher
doping; the intrinsic width of PPV might therefore be lower than
0.19 eV. At very low doping the influence of the ions is expected
to be small. Here, around $g \sim 10^{-4}{\rm
states/(eV\,monomer)}$, the data allow for fitting to a Gaussian
with $\sigma_{\rm d}$ = 0.11 eV (horizontally shaded area B in
Fig.~\ref{densitE}b), as reported in Refs.~\onlinecite{Martens00}
and \onlinecite{Tanase03}. Note, see Fig.~\ref{densitE}b, that the
Gaussian distribution in this doping regime is a tail effect,
which is not representative for the main distribution.

On inspection of the DOS it is clear that the assignment of the
HOMO level at -5.2~eV \cite{vanDuren03,Hulea03} (horizontal dashed
line in Fig~\ref{densitE}), as determined by cyclic voltammetry
(location at 10\% of the peak value), is relatively arbitrary. The
value of -5.2~eV corresponds to the point at which a strong
increase in DOS is observed. However, many states are already
available for values of $E$ up to -4.5 eV. This means that in LED
applications a deviation of the ''HOMO level'' of 0.7~eV from the
cyclic voltammetric value is possible. This has to be included in
quantitative descriptions of charge injection and carrier mobility
in devices \cite{Arkhipov03}.

\begin{figure}[t]
\begin{center}
\includegraphics[width=8cm]{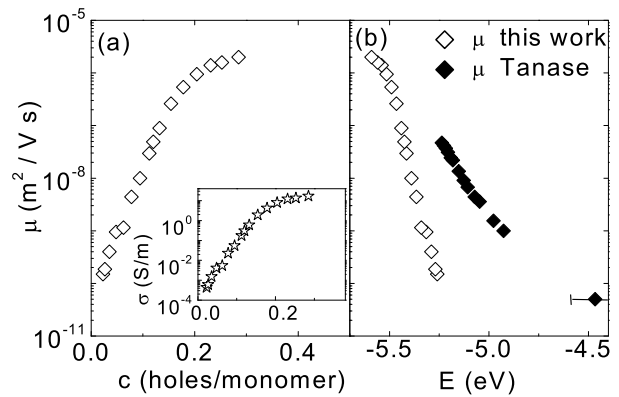}
\end{center}
\noindent{\caption{(a)The hole mobility ($\mu$) calculated from
the conductivity ($\sigma$- the inset) as function of doping.
Because the measurements had to be performed in a two-point
contact configuration, the flattening of $\sigma$ at high $c$
might be caused by contact resistances. Hence, actual values of
$\sigma$ or $\mu$ might be even higher. (b) The dependence of hole
mobility on energy or electrochemical potential $\tilde{\mu_{e}}$.
At lower doping, corresponding to energies $E
>$-5.2~eV, we have included $\mu$ (closed symbols) determined in
a solid-state FET and LED by Tanase {\it et al.} \cite{Tanase03}.
The deviations are discussed in the text. Both plots are on a
logarithmic-linear scale.} \label{cond}}
\end{figure}

When a small voltage difference is applied between the
source-drain contacts, the conductivity $\sigma$ of the PPV layer
can be measured as a function of doping regulated by
$\tilde{\mu_e}$, see Fig.~\ref{cond}. The conductivity, which was
checked to be ohmic up to the highest field of ${\rm 10^4\, V/m}$,
increases by five orders of magnitude, as the doping is increased
from about 0.02 to 0.30 holes per monomer. A similar strong
increase was previously reported for FeCl$_3$ doped PPV
\cite{Martens03}. The values of $\sigma$ are calculated with a
constant thickness of PPV, not taking into account effects of
swelling. Because the measurements had to be performed in a
two-point contact configuration, the flattening of $\sigma$ at
high $c$ might be caused by contact resistances, which means that
actual values of $\sigma$ or $\mu$ might be even higher. From the
measured conductivity the hole mobility ($\mu$) can be determined
via the relation: $\sigma = N p_t e \mu$ with $p_t$ the number of
holes per monomer participating in transport and $e$ the
elementary charge. The states involved in hopping are those within
$k_{\rm B}T$ of the chemical potential \cite{notediffusion},
$p_t(\tilde{\mu_{e}})\sim \int_{\tilde{\mu_{e}}-k_{\rm
B}T/2}^{\tilde{\mu_{e}}+k_{\rm B}T/2}g(E)dE$. The results,
presented in Fig.~\ref{cond}, confirm that the mobility is
concentration, or better, energy dependent
\cite{Martens03,Tanase03}. For a doping level of 0.3 a mobility of
$2\times 10^{-6}$m$^2$/V s is obtained, more than a factor of ten
higher than the highest value found by Tanase {\it et al.}
\cite{Tanase03} for comparable field values, but at 0.02 holes per
monomer. The lower mobilities obtained in the EGT compared to the
solid-state FET at the same doping levels are likely due to the
presence of the anions that tend to localize the mobile holes. In
addition, in the solid-state FET transport takes place primarily
in the first PPV layers close to the interface. The higher
structural order will have a favorable effect on the mobility.

In conclusion, by using an EGT the DOS of a polymer can be
determined over more than one eV or a doping range extending over
four orders of magnitude. A comparison with the DOS obtained with
a solid-state FET \cite{Tanase03} reveals that data below 0.1
states $/$ (eV monomer) (doping levels below $10^{-2}$ holes per
monomer) are representative for the intrinsic level distribution
of ${\rm OC_1C_{10}}$-PPV. The shape of the DOS function is
complex. While the core is well described by a Gaussian function
(its width of 0.19 eV has to be considered an upper boundary for
the intrinsic width), the flank decays exponentially and the tail
at very low densities allows a description with another Gaussian.
The hole mobility is found to vary from $10^{-10} {\rm m^2/Vs}$ to
$10^{-6}{\rm m^2/Vs}$, as the chemical potential changes from -5.3
eV to -5.6 eV. We have shown that PPV can be electrochemically
doped with a high degree of control and reversibility. The
accessible energy window is much larger than that of previously
used methods. Thus electrochemical gating appears to be a powerful
method for studying charging and transport properties of
conducting polymers.

We wish to thank Frank Janssen (TUE) for preparing the samples and
Peter Liljeroth (UU) for experimental assistance and we
acknowledge fruitful discussions with Paul Blom and Cristina
Tanase (RUG) about the solid-state FET data. This work forms part
of the research program of the Dutch Polymer Institute (DPI),
project DPI274.


\begin{thebibliography}{10}
\bibitem{noteelectrondope}
{Attempts to dope the PPV with electrons failed because of
irreversible  reactions with the PPV; double valent ions could not
be dissolved.}
\bibitem{noteconversion}
{For conversion of the DOS values vs. doping to an energy-scale
the monomer volume of 0.487 nm$^3$ was calculated from mass and
specific weight.}
\end{thebibliography}
\end{document}